\documentclass{article}
\usepackage{amsmath}

\input{tcilatex}

\begin{document}

\title{The dissipation of the system and the atom in two-photon
Jaynes-Cummings model with degenerate atomic levels}
\author{Y. Q. Guo, L. Zhou, H. S. Song \\
%EndAName
Department of Physics, Dalian University of Technology, \\
Dalian 116024, China}
\maketitle

\begin{abstract}
The method of perturbative expansion of master equation is employed to study
the dissipative properties of system and of atom in the two-photon
Jaynes-Cummings model (JCM) with degenerate atomic levels. The numerical
results show that the degeneracy of atomic levels prolongs the period of
entanglement between the atom and the field. The asymptotic value of atomic
linear entropy is apparently increased by the degeneration. The amplitude of
local entanglement and disentanglement is suppressed. The better the initial
coherence property of the degenerate atom, the larger the coherence loss.
\end{abstract}

\baselineskip=20pt

\textit{PACS: 42.50.Pq, 42.50.-p, 03.67.-a}

\section{Introduction}

A main obstacle of quantum information is the decoherence of qubits. A
quantum system inevitably interacts with its surrounding environment, and
the interaction between the quantum system and its surrounding could lead to
decoherence. Therefore, studying decoherence process \cite{1,2,3} and
solving the problem \cite{4} have attracted much attention. Many authors
stressed that the decoherence is induced by the role of environment
dissipation \cite{1,2,3}.\quad

As a candidate of quantum information, cavity QED has been widely studied in
follow aspects: quantum network \cite{5,6}, producing entanglement\cite{7,10}
and quantum teleportation \cite{8,9,10}, especially in the dissipative
dynamics \cite{2,3,11}. This is because: the dissipation of cavity QED could
be monitored in experiment \cite{2,11}, and this is valuable not only to the
candidate of quantum information but also to the research of decoherence
theory. In Ref. \cite{3}, the dissipative dynamics of the JCM in the
dispersive approximation has been studied, and the influence of dissipation
on the entanglement of the two subsystems was investigated. In Ref. \cite{12}%
, the authors studied the field state dissipative dynamics of two-photon JCM
and showed that the cavity dissipation affected the coherence properties of
the field. In these studies, the atom is a single pure two-level one. While,
in some case, atomic level is degenerate in the projections of the angular
momenta on the quantization axis \cite{13}. In Ref. \cite{14,15}, the
authors studied the case of degenerate atomic levels and showed that the
revival period of the atom population inversion became longer than that of
the original JCM. In Ref. \cite{16}, our group has studied the effect of
degenerate atomic levels on the field state dissipation in two-photon JCM.
The initial degenerate atomic states are found to increase the period of
entanglement of atom and field, and the coherence properties of field is
affected by the reservior qualitatively.

The interaction between an atom and a single-mode field via two-photon
process is a nonlinear one. Giving the explicit solution of the dissipative
dynamics of atom in a nonlinear process is very difficulty. So, the atomic
decoherence property in a two-photon JCM with degenerate atomic levels is
not studied before, despite that the field decoherence property has been
studied \cite{12,16}. However, the atom is the information depositor in a
quantum information system \cite{17}; it is more important to study the
atomic decoherence property than that of the field. In the present paper,
considering degenerate atomic levels and using the method of perturbative
expansion for master equation, we study the dissipative dynamics of system
and of atom in two-photon JCM. We show that the period of entanglement
between atom and field is prolonged explicitly. In case of degenerate atomic
levels, contrasting to the case of two levels atom, the atomic asymptotic
value of linear entropy is increased clearly in a same dissipative cavity.
This asymptotic value is the largest when the atom is initially in an equal
probability superposition degenerate state. Moreover, the numerical results
show that the degenerate atomic levels increase the mixture degree of
nonlinear situation of two-photon JCM.

\section{Perturbative solution of master equation}

Let us consider an atom with degenerated levels interacts with a single mode
electromagnetic field at zero temperature as usual in a high-Q cavity. The
dynamics of the global density operator $\hat{\rho}$ in interaction
representation is governed by the master equation

\begin{equation}
\frac{d}{dt}\hat{\rho}=-i\left[ H_{eff},\hat{\rho}(t)\right] +\kappa (2a\hat{%
\rho}a^{\dagger }-aa^{\dagger }\hat{\rho}-\hat{\rho}aa^{\dagger }),
\end{equation}%
where $\kappa $ is the damping constant of the cavity. The effective
Hamiltonian $H_{eff}$ in two-photon JCM with atomic degeneracy is nonlinear.
Thus Eq. (1) is a nonlinear differential equation. Completely solving the
equation with the usual method used in the study of original JCM is very
difficult. In order to study the dissipative dynamics of two-photon JCM, we
can solve the master equation approximately \cite{18}. The $\hat{\rho}(t)$
can be expanded in powers of $\kappa $ as

\begin{equation}
\hat{\rho}(t)=\hat{\rho}^{0}(t)+\frac{\partial \hat{\rho}(t)}{\partial
\kappa }\kappa +\frac{1}{2}\frac{\partial ^{2}\hat{\rho}(t)}{\partial \kappa
^{2}}\kappa ^{2}+0(\kappa ^{2}).
\end{equation}%
In a high-Q cavity, $\kappa \gg 1$, the effectiveness of the expansion is
assured and the reservation of only fore 3 terms is satisfying.

From Eq. (1), each orders of differential of $\hat{\rho}(t)$ can be
determined by using iterative method from Eq. (1) as%
\begin{equation}
\frac{d}{dt}\hat{\rho}^{(n)}(t)=-i\left[ \hat{H}_{eff},\hat{\rho}^{(n)}(t)%
\right] +(2a\hat{\rho}^{(n-1)}(t)a^{\dagger }-aa^{\dagger }\hat{\rho}%
^{(n-1)}(t)-\hat{\rho}^{(n-1)}(t)aa^{\dagger }),
\end{equation}%
with $n=1,2,3...$,and $\hat{\rho}^{(n)}(t)=\frac{\partial ^{n}\rho (t)}{%
\partial \kappa ^{n}}.$ The solution of Eq. (3) is

\begin{eqnarray}
\hat{\rho}^{(n)}(t) &=&e^{-i\hat{H}_{eff}t}\{\int\nolimits_{0}^{t}dte^{i%
\hat{H}_{eff}t}[2a\hat{\rho}^{(n-1)}(t)a^{\dagger }-aa^{\dagger }\hat{\rho}%
^{(n-1)}(t)-\hat{\rho}^{(n-1)}(t)  \notag \\
&&aa^{\dagger })]e^{-i\hat{H}_{eff}t}\}e^{i\hat{H}_{eff}t}.
\end{eqnarray}%
So, $\hat{\rho}^{(1)}(t)$ can be calculated from an integer of $\hat{\rho}%
^{(0)}(t)$, and $\hat{\rho}^{(2)}(t)$ from $\hat{\rho}^{(1)}(t)$, ... . Thus
we can get arbitrary value of $\hat{\rho}(t)$ at least in principle.

We assume that the initial states of the atom-field system can be written as

\begin{equation}
\Psi (0)=(\sum_{m_{b}}\frac{e}{\sqrt{2J_{b}+1}}\left|
J_{b},m_{b}\right\rangle +\sum_{m_{c}}\frac{f}{\sqrt{2J_{c}+1}}\left|
J_{c},m_{c}\right\rangle )\otimes \left| \alpha \right\rangle ,
\end{equation}%
where $J_{b}$, $J_{c}$ denote the values of the total electronic angular
momenta of resonant levels $b$, $c$ respectively, and $m_{b}$, $m_{c}$
represent their projections on the polarized direction. $\left| \alpha
\right\rangle $ is the coherence field state of cavity; it can be expanded
as $\left| \alpha \right\rangle =\exp (-\frac{\left| \alpha \right| ^{2}}{2}%
)\sum\limits_{n}\frac{\alpha }{\sqrt{n!}}\left| n\right\rangle
=\sum\limits_{n}F_{n}\left| \alpha \right\rangle $. From Eq. (5), we easily
obtain the initial density of the system, i.e. $\hat{\rho}^{(0)}(0)$, and
further obtain the evolution of $\hat{\rho}^{(0)}(0)$. In interaction
representation, the effective Hamiltonian of the two-photon JCM within
rotating-wave approximation including degenerate atomic levels has been
obtained in Ref. \cite{16} as

\begin{equation}
\hat{H}_{eff}=\Omega \left[ (a^{\dagger }a+1)(a^{\dagger
}a+2)R_{b}-(a^{\dagger }a-1)R_{c}\right] ,
\end{equation}
where $R_{\mu }=\sum\limits_{m\mu }\alpha _{m\mu }^{2}(\left| J_{\mu
},m_{\mu }\right\rangle \left\langle J_{\mu },m_{\mu }\right| ),\mu =b,c$. $%
\Omega $ is the Rabi frequency which measures the coupling between atom and
field. $\alpha _{m\mu }$ is defined as $\alpha _{m\mu }=(-1)^{J_{b}-m}\left( 
\begin{array}{lll}
J_{b} & 1 & J_{c} \\ 
-m & 0 & m%
\end{array}%
\right) $.

After some deducing, we obtain the density matrix of the system as

\begin{eqnarray}
\hat{\rho}(t) &=&\sum\limits_{n,n^{\prime },m,m^{\prime }}F_{n}F_{n^{\prime
}}^{\ast }\{\exp \{-i\Omega t[(n+1)(n+2)\alpha _{m}^{2}-(n^{\prime
}+1)(n^{\prime }+2)\alpha _{m^{\prime }}^{2}]\}  \notag \\
&&\frac{e^{2}}{2J_{b}+1}A_{bbnn^{\prime }mm^{\prime }}\left|
n,J_{b},m\right\rangle \left\langle n^{\prime },J_{b},m^{\prime }\right|
+\exp \{i\Omega t[n(n-1)\alpha _{m}^{2}-  \notag \\
&&n^{\prime }(n^{\prime }-1)\alpha _{m^{\prime }}^{2}]\}\frac{f^{2}}{2J_{c}+1%
}A_{ccnn^{\prime }mm^{\prime }}\left| n,J_{c},m\right\rangle \left\langle
n^{\prime },J_{c},m^{\prime }\right| +\exp \{-i\Omega t  \notag \\
&&[(n+1)(n+2)\alpha _{m}^{2}+n^{\prime }(n^{\prime }-1)\alpha _{m^{\prime
}}^{2}]\}\frac{ef}{\sqrt{(2J_{b}+1)(2J_{c}+1)}}B_{bcnn^{\prime }mm^{\prime }}
\notag \\
&&\left| n,J_{b},m\right\rangle \left\langle n^{\prime },J_{c},m^{\prime
}\right| +\exp \{i\Omega t[n(n-1)\alpha _{m}^{2}+(n^{\prime }+1)(n^{\prime
}+2)\alpha _{m^{\prime }}^{2}  \notag \\
&&]\}\frac{ef}{\sqrt{(2J_{b}+1)(2J_{c}+1)}}B_{bcnn^{\prime }mm^{\prime
}}^{\ast }\left| n,J_{c},m\right\rangle \left\langle n^{\prime
},J_{b},m^{\prime }\right| ,
\end{eqnarray}%
where we have set

\begin{eqnarray*}
A_{\mu \mu nn^{\prime }mm^{\prime }} &=&1+\frac{\kappa }{\Omega }[\pm
iNF(n,n^{\prime },m,m^{\prime },t)-(n+n^{\prime })\Omega t]+\frac{1}{2}(%
\frac{\kappa }{\Omega })^{2}\{ \\
&&\frac{N^{2}}{n\alpha _{m}^{2}-n^{\prime }\alpha _{m^{\prime }}^{2}}%
[F(n,n^{\prime },m,m^{\prime },t)-\frac{1}{2}F(n,n^{\prime },m,m^{\prime
},2t)]- \\
&&N[\frac{F(n,n^{\prime },m,m^{\prime },t)}{n\alpha _{m}^{2}-n^{\prime
}\alpha _{m^{\prime }}^{2}}\pm i(n+n^{\prime }+2)F(n,n^{\prime },m,m^{\prime
},t)\Omega t\pm \\
&&\frac{i2\Omega t}{n\alpha _{m}^{2}-n^{\prime }\alpha _{m^{\prime }}^{2}}]+%
\frac{(n+n^{\prime })^{2}}{2}(\Omega t)^{2}\},
\end{eqnarray*}%
\begin{eqnarray*}
B_{bcnn^{\prime }mm^{\prime }} &=&1+\frac{\kappa }{\Omega }[iNE(n,n^{\prime
},m,m^{\prime },t)-(n+n^{\prime })\Omega t]+\frac{1}{2}(\frac{\kappa }{%
\Omega })^{2}\{ \\
&&\frac{N^{2}}{(n+3)\alpha _{m}^{2}+(n^{\prime }+1)\alpha _{m^{\prime }}^{2}}%
[E(n,n^{\prime },m,m^{\prime },t)-E(n+\frac{1}{2}, \\
&&n^{\prime }+\frac{1}{2},m,m^{\prime },2t)]-N[\frac{E(n,n^{\prime
},m,m^{\prime },t)}{(n+2)\alpha _{m}^{2}+n^{\prime }\alpha _{m^{\prime }}^{2}%
}+(n+n^{\prime }+2) \\
&&i\Omega tE(n,n^{\prime },m,m^{\prime },t)+\frac{i2\Omega t}{(n+2)\alpha
_{m}^{2}+n^{\prime }\alpha _{m^{\prime }}^{2}}]+\frac{1}{2}(n+n^{\prime
})^{2}(\Omega t)^{2},
\end{eqnarray*}%
with 
\begin{equation*}
F(n,n^{\prime },m,m^{\prime },t)=\frac{\exp \{-i2\Omega t(n\alpha
_{m}^{2}-n^{\prime }\alpha _{m^{\prime }}^{2})\}-1}{n\alpha
_{m}^{2}-n^{\prime }\alpha _{m^{\prime }}^{2}},
\end{equation*}%
\begin{equation*}
E(n,n^{\prime },m,m^{\prime },t)=\frac{\exp \{-i2\Omega t[(n+2)\alpha
_{m}^{2}+n^{\prime }\alpha _{m^{\prime }}^{2}]\}-1}{(n+2)\alpha
_{m}^{2}+n^{\prime }\alpha _{m^{\prime }}^{2}},
\end{equation*}%
$N=\left| \alpha \right| ^{2}$, and $+$, $(-)$ is chosen from $\pm $ when $%
\mu =b,c$ respectively.

Now,taking the trace of $\hat{\rho}(t)$ on field states variables, we can
obtain the reduced atomic density%
\begin{eqnarray}
\hat{\rho}_{a}(t) &=&\sum\limits_{m,m^{\prime },n,n}F_{n}F_{n}^{\ast }\{%
\frac{e^{2}}{2J_{b}+1}A_{bbnnmm^{\prime }}\left| J_{b},m\right\rangle
\left\langle J_{b},m^{\prime }\right| +\frac{f^{2}}{2J_{c}+1}%
A_{ccnnmm^{\prime }}  \notag \\
&&\left| J_{c},m\right\rangle \left\langle J_{c},m^{\prime }\right| +\exp
\{-i\Omega t[(n+1)(n+2)\alpha _{m}^{2}+n(n-1)\alpha _{m^{\prime }}^{2}]\} 
\notag \\
&&\frac{ef}{\sqrt{(2J_{b}+1)(2J_{c}+1)}}B_{bcnnmm^{\prime }}\left|
J_{b},m\right\rangle \left\langle J_{c},m^{\prime }\right| +\exp \{i\Omega
t[n(n-1)  \notag \\
&&\alpha _{m}^{2}+(n+1)(n+2)\alpha _{m^{\prime }}^{2}]\}\frac{ef}{\sqrt{%
(2J_{b}+1)(2J_{c}+1)}}B_{bcnnmm^{\prime }}^{\ast }\left| J_{c},m\right\rangle
\notag \\
&&\left\langle J_{b},m^{\prime }\right| .
\end{eqnarray}%
The linear entropy is defined as%
\begin{equation}
S=1-Tr(\hat{\rho}^{2}),
\end{equation}%
which can be used to measure the coherence lose and the purity degree of the
state. For a pure state $S=0$, otherwise, $S>0$, corresponding to a mixture.

Substitute $\hat{\rho}(t)$, $\hat{\rho}_{a}(t)$ into Eq. (9). We get the
linear entropy of system $S$ and of atom $S_{da}$. Here, we use letter $d$
to denote atomic linear entropy in JCM in case of degenerate atomic levels.
The linear entropy $S_{da}$ then depends on a set of parameters $(e,f,\left|
\alpha \right| ^{2},J_{b},J_{c},\frac{\kappa }{\Omega },\Omega t)$. In next
section, typical values of the parameters are given to discuss the behavior
of $S$, $S_{da}$, and from the appearance of the linear entropy, the
decoherence of the system and the atom and the disentanglement of atom and
field are studied.

\section{Results and Discussion}

In experiment \cite{19}, the resonant atomic levels $b,c$ were usually the
Rydberg states of the atom with angular momenta $J_{b}=\frac{3}{2},J_{c}=%
\frac{3}{2}$ or $J_{c}=\frac{5}{2}$. By the support of these experiment
data, we choose $J_{b}=J_{c}=\frac{3}{2}$, which emerges $\alpha _{\frac{1}{2%
}}=\alpha _{-\frac{1}{2}}=\frac{1}{2\sqrt{15}}$, $\alpha _{\frac{3}{2}%
}=\alpha _{-\frac{3}{2}}=\frac{3}{2\sqrt{15}}$.

With same set of parameters, we plot the evolution of atomic linear entropy
with and without degeneration to study the effect of atomic degeneracy on
the dissipative dynamics of atom in Fig.1. 

Without degeneration, the atomic is a pure two-level one. Under this case,
the linear entropy has been discussed in Ref. \cite{20}. We directly employ
the result in that paper to show the effect of the atomic degeneracy. In the
process of atomic linear entropy evolution, Fig.1 show that the behavior of $%
S_{a}(t)$ (atomic linear entropy in non-degenerate case) and $S_{da\text{ }}$%
both present local maximum and minimum. This corresponds to the entanglement
and disentanglement respectively of atom and field. In Ref. \cite{3}, the
authors have shown that disentanglement took place at instants $t_{d}=\frac{%
n\pi }{\omega }$. Here, contrasting to the case of non-degeneration \cite{20}%
, the period of entanglement and disentanglement in case of degenerate
atomic levels is obviously prolonged. And the amplitude of local
entanglement and disentanglement is suppressed. We can also observe that the
two asymptotic values are obviously different, i.e. $S_{da}(\infty
)>S_{a}(\infty )$. The linear entropy is seen as a measure of the coherence
loss. If the atom is initially in a degenerate state written in Eq. (5), the
coherence property is better than that of without degeneracy. When coherence
loss finished completely, the better the initial coherence property the
bigger the coherence loss. We will show the effect of initial coherence
property on the coherence loss in Fig.3. Through comparing the atomic linear
entropy with and without degeneracy, we find that the degeneracy increases
the maximum of the asymptotic value of the linear entropy. Thus, the
degeneracy of the atom on one hand increase the period of entanglement on
the other hand increases the atomic maximum coherence loss. The result has
not been given before.

The dependence of the linear entropy on the photon number is shown in Fig.2.
It is clear that the period of entanglement does not depend on the average
photon number $\left| \alpha \right| ^{2}$. And obviously, the larger the
average photon number , the larger the asymptotic value of the curves. The
similar results have been shown in Ref. \cite{3}. In that paper, the
asymptotic values of the linear entropy never exceed $\frac{1}{2}$, while
here, the asymptotic values can do. The average photon number $\left| \alpha
\right| ^{2}$ is a measure of its classicality. The entanglement of atom and
field proportionally contributes to $S_{da}$. Therefore, the larger the $%
\left| \alpha \right| ^{2}$ is the more rapidly the atom and the system lose
their coherence. These properties are also observed in Ref. \cite{3,15}.

Now, we take different initial atomic degenerate states to show their effect
on the disentanglement and decoherence of atom. In Fig.3, when the atom is
initially in degenerate state with $e=f=1/\sqrt{2}$, the asymptotic value is
the largest. In this case, the initial coherence of atom is the best; when
dissipation finally finishes, the coherence loss is the most and the linear
entropy is the largest. Fig.3 also shows that the amplitude of local maximum
and minimum of the curves is the largest in this case. Because of the best
initial coherence property of atom, the atom would best entangle with field.
So, the better the initial atomic coherence the larger the amplitude of
local maximum and minimum. Therefore, the better the initial atomic
coherence properties, the larger the asymptotic value of $S_{da\text{ }}$and
the larger the amplitude of local maximum and minimum of $S_{da\text{ }}$.

\section{Conclusion}

We have studied the dissipative dynamics of system and atom in two-photon
JCM with degenerate atomic levels by employing the perturbtive expansion of
master equation. The results show that the degeneracy prolongs the period of
entanglement and disentanglement of atom and field qualitatively. This
provides us a way in preparing a long time surviving entangled quantum
states needed in experiment \cite{21,22}. The degeneracy also increases the
asymptotic value of the linear entropy. This means the coherence loss is
increased by the degeneracy. Comparing to the non-degenerate case, the
amplitude of local maximum and minimum of the linear entropy is suppressed
due to the degeneracy. The more average photon number affects the cavity
dissipation by increasing the asymptotic value of the linear entropy of the
atom and the system and making them lose their coherence more rapidly than
usual. The better the initial coherence property of the atom the larger the
asymptotic value of the linear entropy, i.e. the larger the coherence loss.

{\Huge Figure Captions:}

{\small Fig.1 Linear entropy of atom as a function of }$\Omega t${\small , }$%
S_{da}${\small \ with degenerate atomic levels (solid line), }$S_{a}${\small %
\ with two levels atomic state (dotted line), where }$e=f=1/\sqrt{2}${\small %
,}$\kappa /\Omega =0.02${\small ,}$\left| \alpha \right| ^{2}=0.5${\small .}

{\small Fig.2a Linear entropy of the system (atom+field) as a function of }$%
\Omega t${\small , where }$e=f=1/\sqrt{2}${\small ,}$\kappa /\Omega =0.01$%
{\small ,solid line with }$\left| \alpha \right| ^{2}=1${\small , dotted
line with }$\left| \alpha \right| ^{2}=0.5${\small .}

{\small Fig.2b Linear entropy of the atom as a function of }$\Omega t$%
{\small , where }$e=f=1/\sqrt{2}${\small , }$\kappa /\Omega =0.02${\small ,
solid line with }$\left| \alpha \right| ^{2}=1${\small , dotted line with }$%
\left| \alpha \right| ^{2}=0.5${\small .}

{\small Fig.3 Linear entropy of atom as a function of }$\Omega t${\small ,
Where }$\kappa /\Omega =0.02${\small , }$\left| \alpha \right| ^{2}=0.5$%
{\small , solid line with }$e=f=1/\sqrt{2}${\small ,dotted line with }$e=0.4$%
{\small .}

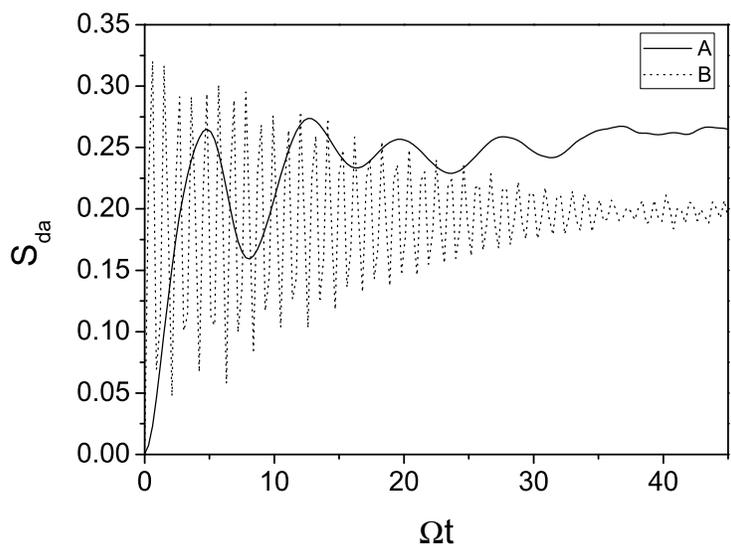
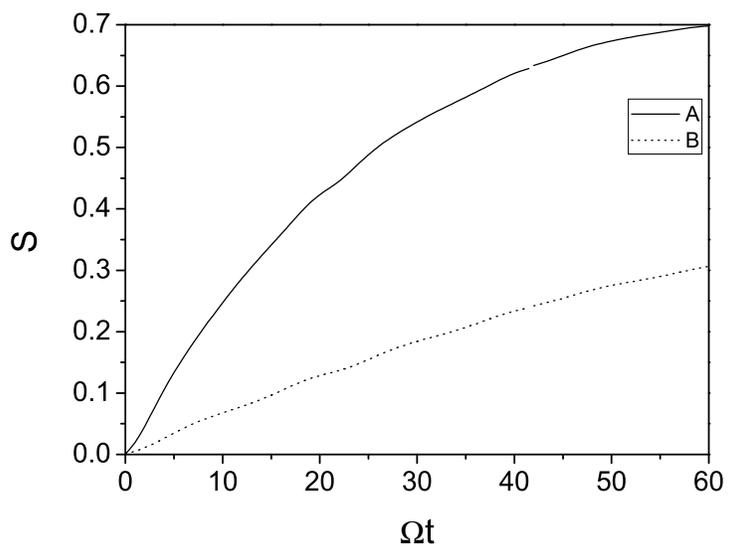
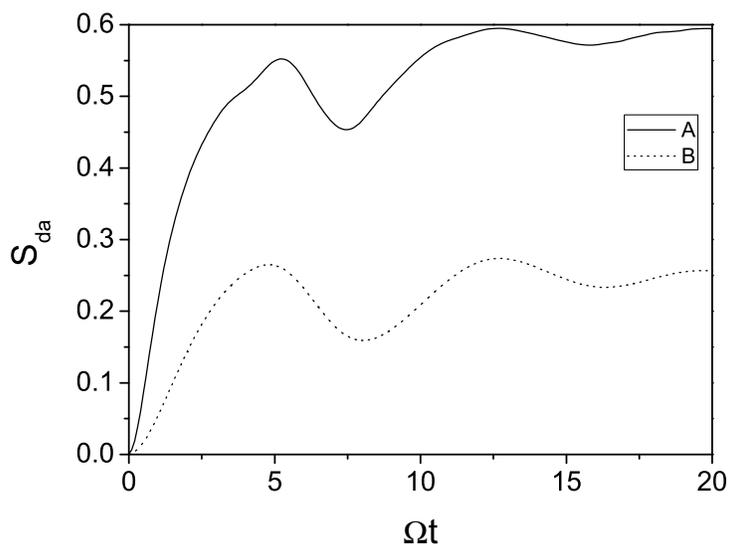
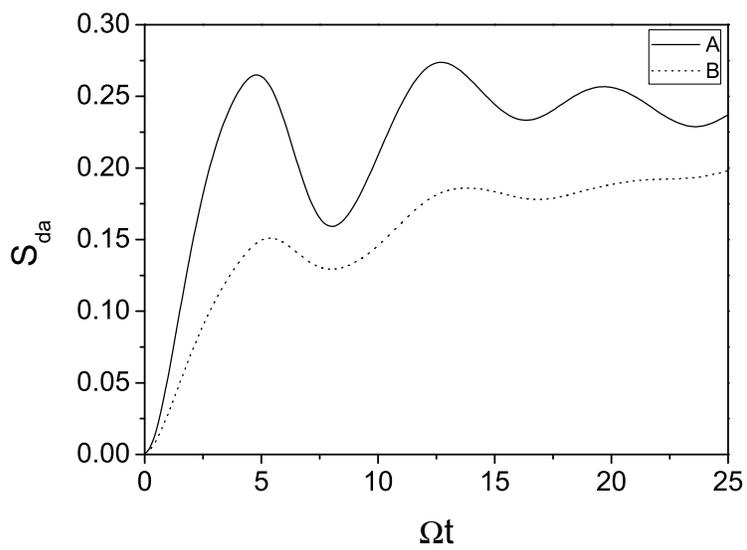
\end{document}